\documentclass[10pt,conference]{IEEEtran}
\IEEEoverridecommandlockouts
\usepackage{cite}
\usepackage{amsmath,amssymb,amsfonts}
\usepackage{algorithmic}
\usepackage{graphicx}
\usepackage{textcomp}
\usepackage{xcolor}
\usepackage{booktabs}
\usepackage{multirow}
\usepackage{url}
\usepackage{tcolorbox}
\def\BibTeX{{\rm B\kern-.05em{\sc i\kern-.025em b}\kern-.08em
    T\kern-.1667em\lower.7ex\hbox{E}\kern-.125emX}}

\makeatletter
\newcommand{\linebreakand}{%
  \end{@IEEEauthorhalign}
  \hfill\mbox{}\par
  \mbox{}\hfill
  \begin{@IEEEauthorhalign}
}
\makeatother

\begin{document}

\title{From Failing to Passing: Evolving Natural Language Prompt Optimization Rules for LLM Code Generation}

\author{
\IEEEauthorblockN{Amal Akli}
\IEEEauthorblockA{
\textit{University of Luxembourg}\\
Luxembourg \\
amal.akli@uni.lu}
\and
\IEEEauthorblockN{Melissa Akli}
\IEEEauthorblockA{
\textit{University of Toulouse}\\
France \\
melissa.akli@utoulouse.fr}
\and
\IEEEauthorblockN{Cedric Richter}
\IEEEauthorblockA{
\textit{University of Luxembourg}\\
Luxembourg \\
cedric.richter@uni.lu}
\linebreakand
\IEEEauthorblockN{Mike Papadakis}
\IEEEauthorblockA{
\textit{University of Luxembourg}\\
Luxembourg \\
michail.papadakis@uni.lu}
\and
\IEEEauthorblockN{Yves Le Traon}
\IEEEauthorblockA{
\textit{University of Luxembourg}\\
Luxembourg \\
Yves.LeTraon@uni.lu}
}
\maketitle

\begin{abstract}
Large language models are known to be sensitive to prompt formulation. Even minor variations in wording can substantially degrade performance. This sensitivity reveals an opportunity: \textit{if prompt phrasing can harm performance, can it be used to improve it?} To investigate this question, we introduce a search-based approach that identifies and evolves a set of natural language transformation rules with strong downstream effects on coding performance. We then propose \textsc{DualFix}, a staged repair pipeline that combines the evolved transformation rules with execution-feedback repair, addressing both specification-level and implementation-level failures. A key strength of our approach lies in its generality: the evolved rules are error-agnostic, reusable across problems, and transferable across models. We evaluate \textsc{DualFix} against execution-feedback repair baselines across three models on two challenging benchmarks, LiveCodeBench and APPS. Our results show that the evolved transformations fix from \emph{10-30\%} of failing cases, including \emph{12--17\%} of failures that execution-based repair alone cannot resolve. Overall, DualFix recovers up to 30\% of baseline failures and fixes 3–5 times more failing cases than Self-Fix across all evaluated settings. Furthermore, we also show that rules evolved on one model transfer zero-shot to other models, outperforming execution-feedback repair without any re-optimization.
\end{abstract}


\section{Introduction}
Large language models (LLMs) have become increasingly capable in code generation tasks, achieving strong performance on benchmarks such as HumanEval~\cite{chen2021evaluating}, APPS~\cite{hendrycks2021apps}, and LiveCodeBench~\cite{jain2025livecodebench}. Despite these advances, the generated code is often incorrect, as it may fail to fully capture or adhere to the user's intent~\cite{austin2021program}. Consequently, users must manually test and validate the produced code to ensure that the generated code meets its expectations. This raises an important question: \textit{how can we effectively and efficiently repair faults in LLM-generated code?}

A key challenge in repairing LLM-generated code is that existing approaches~\cite{DBLP:conf/iclr/ChenLSZ24,DBLP:conf/nips/ShinnCGNY23,madaan2023selfrefine} are often ineffective at fixing errors when relying on the original prompts and test failure signals. As a result, users must engage in multiple iterations of ad-hoc prompt rewriting to obtain acceptable solutions. Unfortunately, the question of automatic repair falls short given the inability of the LLMs to extrapolate correct implementations for challenging cases~\cite{olausson2024selfrepair}. Indeed, our results indicate that performing up to three repair iterations based on the original prompts and test-execution (failure) feedback resolves just 10--19\% of the failing cases.

This suggests that if we can develop generic and effective methods to transform failing code into passing solutions without human intervention, we could significantly mitigate the practical limitations of automated code generation. Such approaches would enable generated code to more reliably conform to both the intended task specification and associated tests, thereby strengthening the effectiveness and applicability of so-called ``vibe coding.''

Existing work on prompt optimization, such as Chain-of-Thought prompting~\cite{wei2022chainofthought}, MIPRO~\cite{opsahlong2024mipro}, and GEPA~\cite{agrawal2025gepa}, has primarily been developed in the context of system-level prompting, where general instructions are optimized to guide model reasoning across tasks. While these approaches have proven effective, they typically require curated examples and validation procedures, making them costly to deploy. Moreover, they often involve evolving a dedicated system prompt for each task, which limits scalability. Crucially, such methods operate at a global prompt level and largely overlook the fine-grained details of individual problem descriptions, leaving task-specific formulation underexplored.

To improve task-specific formulations and increase the likelihood that generated code satisfies the intended requirements, we aim to reformulate problem descriptions in two complementary ways: (a) \textit{rule-based rewriting} - by leveraging, or deliberately avoiding, empirically identified sensitive wording, i.e., phrasing that significantly influences LLM behavior, and (b) \textit{error-based rewriting }  - by integrating test failure signal in a coherent and natural manner by appropriately prompting an LLM to rewrite the task description. 

Both strategies aim to refine the wording and clarify the existing information, thereby better aligning problem statements with the inductive biases of the underlying models and improving the passing rate of downstream code generation. We operationalize these strategies through a staged pipeline, \emph{DualFix}, which first applies error-based rewriting and, when execution feedback alone is insufficient, falls back to rule-based rewriting.

Conceptually, our approach is analogous to compiler optimization: just as compilers transform code to improve execution efficiency, DualFix operates at the level of natural language, optimizing task descriptions to enhance downstream model performance. Unlike traditional compiler optimizations, however, our transformations target the prompt itself, aiming to reduce the likelihood of incorrect code generation by improving how the task is communicated to the model.

To transform sensitive wording, i.e., phrasing that strongly influences LLM code generation performance,  we adopt a search-based approach \cite{HarmanJ01,HarmanMZ12} that identifies and evolves natural language transformation rules. The learned rules are derived from a set of example tasks and a target model, yet remain error-agnostic, reusable across different problems, and independent of any specific code generation model. As a result, once learned, these rules can be applied without requiring additional examples or further optimization. This enables a lightweight and scalable mechanism for improving code generation through prompt rewriting alone, without modifying the underlying model or relying on costly task-specific tuning.

We empirically evaluate our approach on the APPS and LiveCodeBench benchmarks. We compare against the current state of practice, namely, direct code correction based on the addition of execution-failure feedback (self-fix)~\cite{DBLP:conf/iclr/ChenLSZ24,olausson2024selfrepair}, as well as through ablation studies of our two main components: (i) the natural language transformation rules and (ii) the LLM-based integration of execution failure signals into prompt descriptions. Our results show that our approach turns to passing up to 30\% of baseline failures on LiveCodeBench and up to 21.3\% on APPS, fixing 3--5$\times$ more problems than Self-Fix by leveraging signal by only 1 failing test.

Our ablation study further demonstrates that the evolved transformation rules alone makes passing up to 21.6\% of failures on LiveCodeBench and 13.8\% on APPS, including cases that execution-based repair is unable to resolve in any of the evaluated settings. Similarly, incorporating failure signals through LLM-based prompt rewriting turns to passing up to 19.5\% of failures on LiveCodeBench and 15.0\% on APPS, compared to only 7.9\% and 5.9\% achieved by naively appending failure messages (Self-Fix). Furthermore, these gains are consistent across all evaluated models. Finally, our analysis of cross-model transferability shows that rules evolved on one model transfer zero-shot to an unseen model, turning passing 45.8\% of failures on LCB and outperforming all error-based baselines without any re-optimization.

Interestingly, our rule-based optimization approach may extend beyond the domain of code generation. In particular, model-specific optimization of the language used can be both feasible and effective in other domains, such as mathematical problem solving and general reasoning tasks. For instance, such an approach could support beginners and non-experts by providing automatically crafted transformation rules that enhance LLM performance. 

The main contributions of this paper are:
\begin{itemize}
\item We propose DualFix, a staged pipeline that composes specification rewriting with execution-feedback repair, and show that it consistently outperforms either strategy alone across models and benchmarks.

\item We introduce a search-based framework that evolves reusable, error-agnostic transformation rules to rewrite problem statements prior to code generation.

\item We show that the evolved rules transfer zero-shot to other models, demonstrating that they capture general patterns rather than model-specific artifacts.

\end{itemize}
\section{Related work}

\subsection{Prompt Sensitivity of Large Language Models}
Large language models are highly sensitive to prompt design~\cite{DBLP:conf/iclr/Sclar0TS24, DBLP:conf/iclr/SunSW24, DBLP:conf/emnlp/IsmithdeenKK25, DBLP:journals/tacl/MizrahiKMDSS24,DBLP:conf/emnlp/ZhuoZFDL024, DBLP:conf/emnlp/ChatterjeeRB024}. Even small perturbations such as syntactical changes to the prompt~\cite{DBLP:conf/iclr/Sclar0TS24, DBLP:conf/emnlp/ZhuoZFDL024}, rephrasing of instructions~\cite{DBLP:conf/iclr/SunSW24, DBLP:conf/emnlp/ChatterjeeRB024}, and other linguistic modifications~\cite{DBLP:conf/emnlp/IsmithdeenKK25,fagadau2024copilot,ma2025promptstability,DBLP:conf/acl/VoronovWR24}, influence whether the model is able to answer a question correctly. These findings are not limited to a specific task, but also translate into the coding domain. ReCode~\cite{wang2023recode} showed that perturbations of signatures and docstrings can impact the ability of language model to complete an implementation. Chen et al.~\cite{DBLP:journals/tosem/ChenLHX26} showed that small changes of natural language requirements leads to the generation of incorrect code. Rabbi et al.~\cite{DBLP:journals/ese/RabbiDY26,zi2025partialorder} shows that this observation generalizes to various programming tasks in different programming languages. 

Beyond surface-level perturbations, Larbi et al.~\cite{larbi2025promptsgowrong} demonstrated that ambiguous, contradictory, or incomplete task descriptions cause pass@1 drops of 20--40\%, with 60--90\% of the syntactically valid code being underspecified. Akli et al.~\cite{akli2026defective} further showed that the severity of such defects depends on the benchmark, with harder task descriptions like LiveCodeBench\cite{jain2025livecodebench} providing greater resilience.

In this paper we view this sensitivity to prompt phrasing as an opportunity rather than a limitation. The same perturbations that can cause a model to generate incorrect code may equally steer it towards a correct solution \cite{akli2026underspec}. We therefore aim at improving LLM's performance through systematic natural language transformations that exploit model's biases and imperfections.  

\subsection{Program Repair from Execution-based Feedback}
Program repair has become a central component of LLM-based code generation~\cite{DBLP:conf/icse/XiaWZ23, DBLP:conf/iclr/ChenLSZ24, DBLP:conf/nips/ShinnCGNY23, DBLP:conf/issta/Xia024}. Rather than treating code generation as a one-shot problem, a growing body of work investigates iterative repair loops in which a generated program is executed against a test suite and the resulting feedback is fed back to the model to guide correction.  Chen et al.~\cite{DBLP:conf/iclr/ChenLSZ24} introduced Self-Debugging, which uses execution traces to iteratively repair model-generated code. Shinn et al.~\cite{DBLP:conf/nips/ShinnCGNY23} proposed to accumulate execution outcomes as verbal feedback. Xia et al.~\cite{DBLP:conf/issta/Xia024,jia2025specfix} has found that LLMs are capable to fix a variety of bugs from execution-based feedback. 

All of these approaches have a common assumption that test failures stem from an incorrect translation of the specification. Our key insight in this paper is that iterative repair in LLM-based code generation is a \emph{dual-level problem}: test failures may stem from either (1) an incorrect implementation of the specification, or (2) a suboptimal specification that leads the model to an interpretation diverging from the developer's intent. While existing approaches address only the first level, this paper exploits both, treating specification perturbation as a complementary repair strategy alongside code-level fixes.

\subsection{Prompt Optimization}
As the effectiveness of large language models is highly dependent on the quality of the prompt, there is growing interest in automated approaches for prompt optimization. While early works focused on gradient-based methods~\cite{DBLP:conf/acl/LiL20} to optimize prompt embeddings, these approaches are often computationally expensive and inapplicable to closed-source models. Therefore, gradient-free methods~\cite{DBLP:conf/iclr/ZhouMHPPCB23, DBLP:journals/corr/abs-2309-03409, opsahlong2024mipro, agrawal2025gepa} have been proposed as an alternative. Existing works, such as APE~\cite{DBLP:conf/iclr/ZhouMHPPCB23}, OPRO~\cite{DBLP:journals/corr/abs-2309-03409}, MIPRO~\cite{opsahlong2024mipro}, and GEPA~\cite{agrawal2025gepa}, iteratively generate prompt candidates through LLMs and refine them according to task performance. 

Overall, all these approaches optimize a single prompt template intended to generalize across every task instance. Thus, they can discover general ``systemic'' improvements of the prompt, such as the inclusion of imports that are typically useful for all coding tasks, but do not optimize specific descriptions or correct the phrasing of individual task specifications. Nevertheless, these approaches are not used to turn failing into passing code generation approaches as they are applied to the system prompts before any code generation. In this work, we aim to optimize a set of specification-level transformation that can be applied to individual instances, enabling the model to generate better code for tasks where the original specification led to failing code. 
\section{Proposed Approach}
\subsection{Overview}

\begin{figure}
    \centering
    \includegraphics[width=0.98\linewidth]{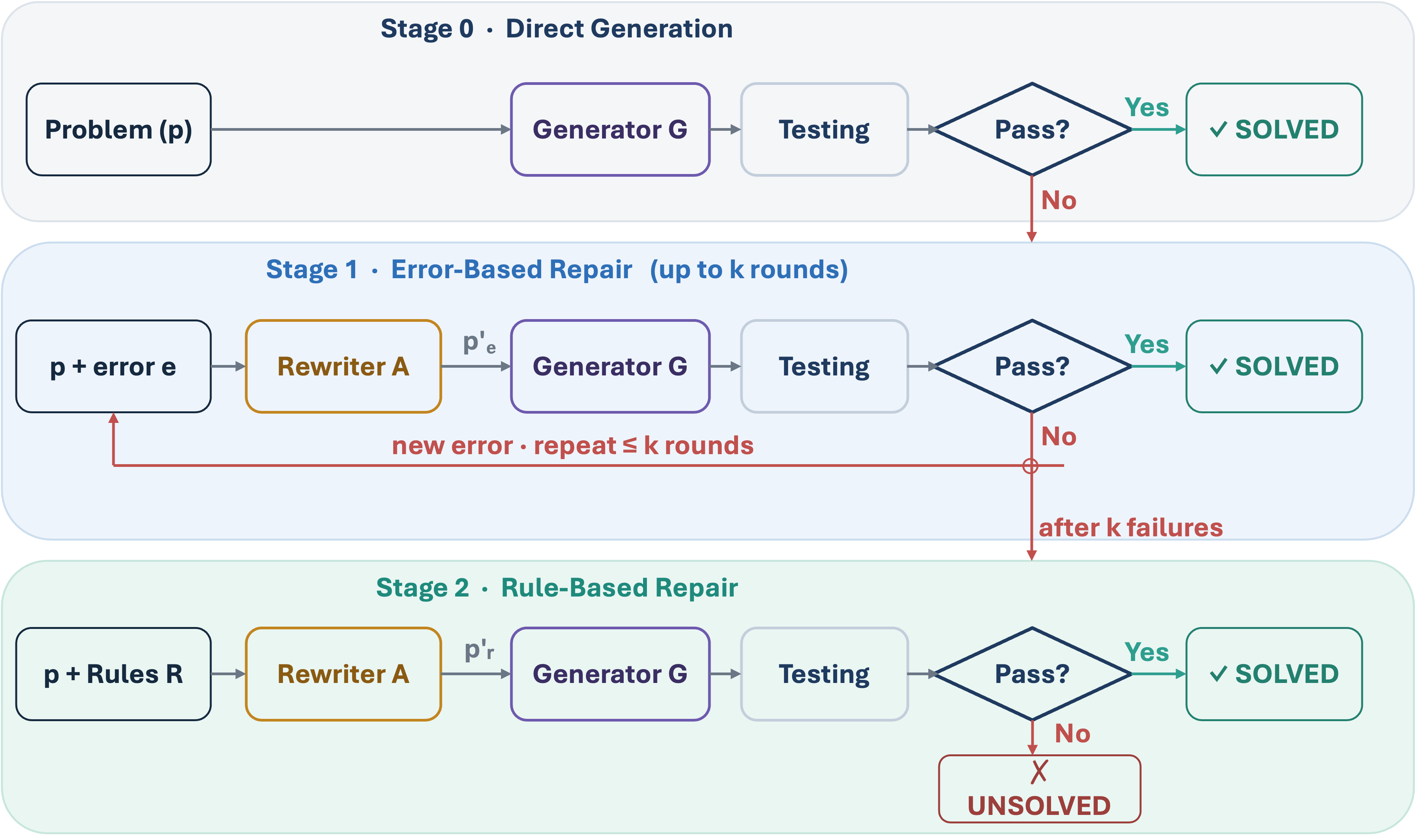}
    \caption{The DualFix inference pipeline: Error-based repair is attempted first; if the problem remains unsolved, rule-based rewriting is applied. It must be noted that DualFix only requires one failing test to operate.}
    \label{fig:dualfix}
\end{figure}

We propose DualFix, an approach that optimizes a set of natural language transformation rules $R$ that rewrite a problem statement in a way that is better aligned with the target LLM $G$, before it is used for code-generation. $G$ is kept frozen: we never change its weights, only the task description that $G$ reads is changed. By transforming the task description, $G$ becomes more likely to produce correct code, i.e., code that passes the held-out tests.

Our approach operates in two phases. In an offline phase (\emph{rule evolution phase}) that happens during ``training time'', our approach uses genetic search to discover sensitive phrases that when changed impacts the execution outcome, turning failed code generation attempts into passing ones. 
Our objective function is thus determined by the pass-fail behaviour of the generated code and we seek to maximize the generation of passing code. 

In the second phase, the \emph{inference phase}, the evolved rules are used to rewrite the task description whenever the code generator fails with the original description, supporting iterative program fixing with an LLM. A staged pipeline that we call DualFix is formed by putting together these two components, i.e., an approach that rewrites task description first by using error-based feedback, followed then by a rule-based rewriting that applies our transformation rule set.

\subsection{Evolutionary Search: RuleEvol}

\begin{figure*}
    \centering
    \includegraphics[width=0.9\linewidth]{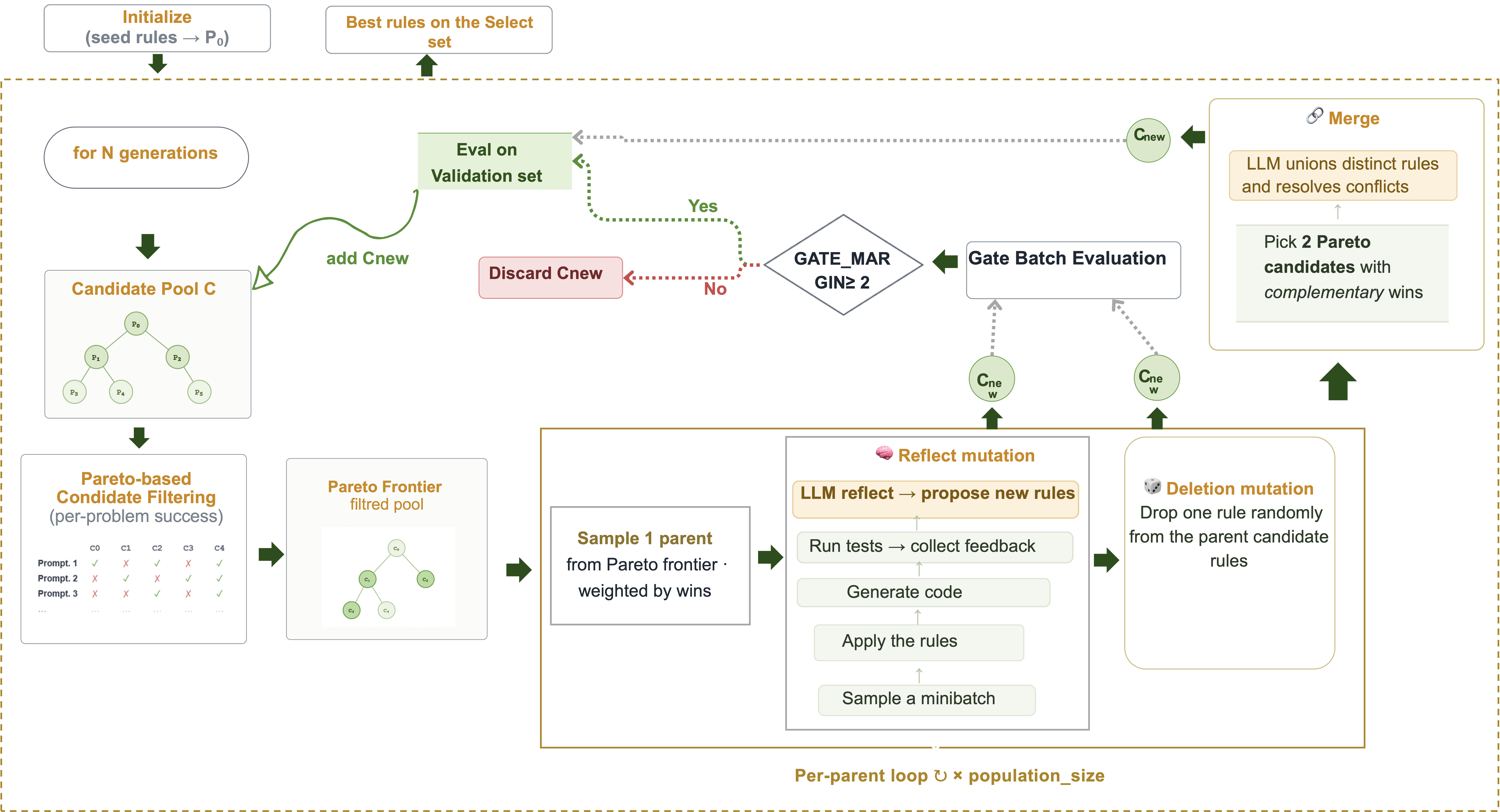}
    \caption{Overview of RuleEvol. A metaheuristic search optimization loop that evolves transformation rule sets over $N$ generations. Each generation samples a parent from the Pareto frontier, applies mutation operators (reflect, delete, merge), evaluates the resulting child on a gate batch, and admits it to the candidate pool only if it improves over its parent. The best rule set is selected on a validation set and is evaluated on a held-out test set after all generations complete.}
    \label{fig:ruleEvol}
\end{figure*}

RuleEvol uses a search-based approach \cite{HarmanJ01,HarmanMZ12} to discover and evolve effective transformation rule sets. The search-based methodology is an adaptation of the GEPA \cite{agrawal2025gepa} framework and involves the following three components (llm requests):

\begin{itemize}
  \item \textbf{Rewriter} $A$, which applies the transformation rules $R$ to an original task description and returns the rewritten description;
  \item \textbf{Generator} $G$, the target code-generation model, which generates code based on the given transformed description;
  \item \textbf{Mutator} $M$, which is used to evolve transformation rules based on feedback from (failing) executions.
\end{itemize}

\subsubsection{Population of Transformation Rules}
\label{sec:rule-format}

Each candidate in the population is a \emph{set of transformation rules} $R$, formatted as a short numbered list of rule definitions. Each rule definition is written in a single \texttt{IF--THEN--EXCEPT} form:

\begin{itemize}
  \item the \textbf{IF} part names a recurring surface-level pattern in the text
        (a kind of notation, phrasing, or layout);
  \item the \textbf{THEN} part defines one small, localized edit that is applied the same way to every match;
  \item the \textbf{EXCEPT} part defines narrow cases where the edit would change
        the meaning and so must be skipped.
\end{itemize}

\noindent \emph{Example rule:}
\begin{quote}\small
\textit{IF} constraints are written with mathematical symbols
($\leq$, $\geq$, \^{}\,), \textit{THEN} rewrite them in plain English while
keeping the exact comparators and limits
(e.g.\ ``$1 \leq n \leq 10^5$'' $\rightarrow$ ``$n$ is at least 1 and at most
100000'').
\end{quote}

A rule may only restate, reorder, or make information explicit that is already defined in the statement. It never adds new facts, hints, or constraints that would alter the semantics of the task description. In particular, rules should never add new facts, constraints, or hints that narrow down the solution space, and it must keep sample inputs, outputs, and numeric values unchanged.

\subsubsection{Initialization}

The population is initialized with a single candidate rule set $R_0$ containing initial handwritten rules that target common surface patterns. The training problems are then partitioned into three disjoint subsets with a fixed random seed: a \emph{development set} used for reflection mini-batches and gate evaluation during the search, a \emph{validation set} used for Pareto ranking, and a \emph{select set} held out for choosing the final rule set after all generations complete. This three-way split ensures that rule selection is never evaluated on the same problems used to guide the evolutionary search.

\subsubsection{Parent selection}
In each generation, candidates are evaluated against the validation set, forming a per-instance Pareto frontier. Each validation problem is claimed by the candidate that scores highest on it, and a candidate survives on the frontier as long as it solves at least one instance not solved better by any other candidate. Parents are sampled from the Pareto frontier via weighted random selection, where the weight of each candidate is proportional to the number of validation instances it uniquely solves. To promote diversity within a generation, a parent is selected at most once per round; if all frontier members have been used, sampling falls back to the full frontier with replacement.

\subsubsection{Mutation operations}

Given a parent candidate, the mutator $M$ is employed to apply one of the following three mutation operators:

\textit{Reflect mutation.} The parent rule set and its execution results on a development minibatch are passed to the Mutator~$M$. The results are categorized into four groups: \emph{helped} (fail$\to$pass), \emph{regressed} (pass$\to$fail), \emph{still failing}, and \emph{stably passing}. The Mutator is instructed to preserve rules behind helped cases, tighten rules that caused regressions, and propose new rules only for wrong-answer failures (ignoring timeouts and import errors, which are not comprehension problems). Each rule is capped at 45 words, and at most two new rules may be added per mutation to prevent unbounded growth.

\textit{Deletion mutation.} A single rule is removed uniformly at random from the parent rule set. This operator tests whether individual rules contribute positively or introduce regressions, acting as a form of ablation within the search.

\textit{Merge.} Two Pareto-frontier candidates with complementary validation wins are selected. The Mutator unions their rules into a single set, deduplicating overlapping patterns and  resolving contradictions. This operator combines discovered patterns into a more comprehensive candidate. The number of merge invocations is bounded per generation to limit cost.

\subsubsection{Mutation Gating}

Each child is evaluated on the gate batch, fresh problems from the development set (different from the mini-batch and larger to generalize more). A child is accepted only if it solves strictly more problems than its parent (margin $\geq 1$), filtering neutral or regressive mutations early. Candidates that are accepted are included in the population of the next generation.

\subsubsection{ Pareto-Based Evolution}
All these steps form a Pareto-based evolution: Candidates survive the mutation process as long as they uniquely solve at least one problem instance. This preserves diversity among transformation rule sets by retaining candidate rule sets with complementary strengths. The final solution to the genetic search process is the rule set that maximizes performance on the held-out select set.

\subsection{Staged Repair Pipeline: DualFix}

At inference time, DualFix employs the evolved rules in a staged pipeline that combines error-based feedback with rule-based rewriting. Given a problem statement $p$, a generator $G$, the generated code $c = G(p)$ that fails, and an error message $e$, the DualFix pipeline proceeds as follows:

\begin{enumerate}
  \item \textbf{Error-based repair.} DualFix first rewrites the original problem statement $p$ based on the error message $e$, producing a revised statement $p'_e$ conditioned on $e$. The rewritten statement is given to $G$ to generate $c'_e = G(p'_e)$ which is evaluated again against the tests. The error message contains the pass/fail count and only the first failing test case's input–output pair; the rewriter never observes the remaining test cases. This step is repeated up to $k$ trials, each time pairing $p$ with the latest error to avoid drift from successive paraphrases. If it passes, DualFix returns $c'_e$.
  \item \textbf{Rule-based repair.} After unsuccessful repair attempts, DualFix applies the evolved rules $R$ to the \emph{original} statement $p$ to obtain $p'_r$, generates code $c'_r = G(p'_r)$ and returns $c'_r$.
\end{enumerate}

\noindent The ordering reflects the complementary nature of our dual-level approach. Error-based repair is attempted first because it has access to a concrete diagnostic signal and can correct implementation-level misunderstandings directly. Rule-based rewriting is applied as a fallback because it targets specification-level clarity issues, ambiguous notation, implicit conditions, and formatting artifacts that produce no informative error message. Crucially, the rule-based stage always rewrites the original statement $p$, not the error-repaired version, ensuring that the two repair approaches operate independently and their contributions do not interfere.

\section{Experimental setup }

\subsection{Research questions}

\paragraph{ \textbf{RQ1. Does DualFix outperform execution-feedback repair in turning the code that initially failed to passing?}}

We aim to evaluate whether DualFix outperforms an LLM purely guided by execution-based feedback. 

\paragraph{\textbf{RQ2. What is the individual contribution of each repair channel to DualFix?}}

We aim to isolate the effect of each component. We evaluate 
rule-based and error-based repair independently and analyze 
the overlap between their fixed sets.

\paragraph{\textbf{RQ3.  Do error-agnostic rules learned on one model transfer to a different code-generation model?}}

We aim to assess generalization. We apply rules evolved for one model directly to another code generation model without any further optimizations.

\subsection{Benchmarks}

We evaluate on two benchmarks that cover distinct problem formats and difficulty ranges.

\subsubsection{LiveCodeBench (LCB) \cite{jain2025livecodebench}}
We use the release-v6 split (1,055 problems) from Codeforces, LeetCode, and AtCoder, mixing functional and stdin/stdout formats. Its rolling release window mitigates data contamination. We stratify-split by source platform with a fixed seed (42), yielding 755 training and 300 test problems. Within the training set, 100 problems are held out for final rule-set selection, leaving 655 for the evolutionary search.

\subsubsection{APPS \cite{hendrycks2021apps}}
We use the test split of APPS, filtered to the "interview" difficulty tier (1,000 problems), which provides sufficient baseline headroom without falling into pure algorithmic-hardness territory. All problems are stdin/stdout format. We apply the same split: 700 training (100 reserved for selection) and 300 test. Test cases are capped at 10 per problem to remain within the evaluation timeout.

\subsection{Models}

RuleEvol assigns code generation, rule application, and rule mutation to three independent frozen models.

\textit{Generator (G).} We evaluate three code-generation models as targets for specification repair: \textit{Qwen2.5-Coder-7B-Instruct} \cite{hui2024qwen25coder}\footnote{https://huggingface.co/Qwen/Qwen2.5-Coder-7B-Instruct}, \textit{Codestral-22B-v0.1} \footnote{https://huggingface.co/mistralai/Codestral-22B-v0.1}, and \textit{Claude Haiku 4.5} \cite{anthropic2025haiku}. All receive an identical system prompt and use greedy decoding. 

\textit{Rewriter (A),} which performs the rule application by \textit{GPT-4o-mini} \cite{openai2023gpt4} through a structured prompt that checks each rule's condition and outputs the rewritten statement. Separating this role from the generator ensures that the quality of the rewritten document does not conflate with the ability to synthesize the code.

\textit{Mutator (M).} Rule-set mutation is performed by \textit{GPT-5-mini} \cite{openai2025gpt5}. The mutator receives the current rule set along with categorized execution feedback and proposes edits. Using a stronger model for this role improves rule quality at minimal cost, since mutation is invoked only once per parent per generation.

\subsection{Baselines and DualFix (ablation) configurations}

All strategies are applied exclusively to problems that the baseline (unmodified prompt, greedy decoding) fails.

\textit{Self-Fix (Baseline).} The generator receives its own failing code and error message and attempts a corrected solution, with no external model involved. The generator acts as both author and repairer. This baseline isolates the generator's capacity to self-correct without additional information.

\textit{Error-Based.} An external model rewrites the problem statement conditioned on the error message, and the generator produces code from the rewritten prompt. 

\textit{Iterative-Error ($\times 3$).} The error-based strategy extended to three rounds, each time pairing the original statement with the latest error to avoid drift.

\subsubsection{Prompt Templates}

The pipeline uses six prompt templates. We summarize their objectives and key constraints below; full templates are available in the replication package.

\textit{Code generation.} The generator receives a system prompt requiring a complete Python solution with explicit imports inside a fenced code block.

\textit{Rule application.} The Rewriter checks each rule's IF-condition against the problem text and applies every matching rule. It must rewrite only the words affected by matching rules, leave everything else identical, not introduce information absent from the original, preserve all numeric values and sample I/O, and keep the output within 2$\times$ the original length.

\textit{Rule mutation.} The Mutator receives the current rules and categorized feedback (helped, regressed, still-failing, stably passing). It must preserve rules behind helped cases, tighten rules that caused regressions, and propose new rules only for wrong-answer failures (ignoring timeouts and import errors). Each rule is capped at 45 words in IF-THEN EXCEPT form, at most two new rules may be added per mutation, and the total rule set may not exceed 10 rules. Rules must be general and content-free, phrased by category, using no token from any specific problem.

\textit{Error-based rewriting.} The Rewriter receives the problem statement and the error message. It must rewrite the statement to help the model avoid the observed error, without adding constraints, examples, or inputs not in the original.

\textit{Combined rewriting.} The Rewriter receives the problem statement, the error message, and the rule set. It must first apply all matching rules, then clarify parts related to the specific error, under the same constraints as rule application.

\textit{Self-Fix.} The generator receives its own failing code, the problem statement, and the error. It must output only corrected code, preserving structure and imports unless the algorithm is fundamentally wrong.

\subsubsection{Technical Details}

All generators use float16 precision, greedy decoding (temperature~$= 0$), and a maximum output length of 2{,}048 tokens. Qwen2.5-Coder-7B runs on a single GPU with batch size 16; Codestral-22B is distributed across two GPUs with batch size 8. The Rewriter and Mutator also operate at temperature~$0$. The evolutionary search runs for 10 generations with 2 children per parent, a Pareto frontier of size 7, a reflection minibatch of 7 problems, a maximum of 10 merge calls, and a gate batch of 40 fresh problems; a child is accepted only if it solves at least 1 more gate problem than its parent. Rule sets are capped at 10 rules, each at most 45 words, and seeded with 6 hand-written rules. Both benchmarks are stratified-split by source platform with a fixed seed ($s{=}42$), reserving 100 problems for validation and 100 for final rule-set selection. Iterative-Error runs up to 3 rounds, always rewriting from the original statement to avoid drift. We report the\textit{ acceptance rate (percentage of baseline failures fixed)} as the primary metric; generated code executes in a sandboxed subprocess with a 30-second timeout, and a problem counts as solved only when all test cases pass.
\section{Results}

\subsection{RQ1: DualFix performance}

\begin{table*}[t]
\vspace{0.5em}
\centering
\small
\caption{Acceptance rate results on LCB and APPS. All repairs are applied to baseline failures only. Rec.\,rate = Fixed / Failures.
\textsc{DualFix} = one error round + rules.
\textsc{Iterative-DualFix} = three error rounds + rules.}
\label{tab:main_results}
\begin{tabular}{l  r r r r  r r r r}
\toprule
 &
  \multicolumn{4}{c}{\textbf{LCB} (300 problems)} &
  \multicolumn{4}{c}{\textbf{APPS} (300 problems)} \\
\cmidrule(lr){2-5}\cmidrule(lr){6-9}
\textbf{Approach} &
 \textbf{Passing} & \textbf{Failing} & \textbf{Fixed} & \textbf{Rec.\,rate} &
\textbf{Passing} & \textbf{Failing} & \textbf{Fixed} & \textbf{Rec.\,rate} \\

\midrule
\multicolumn{9}{l}{\textit{Qwen2.5-Coder-7B-Instruct}} \\
\midrule
Direct generation
  & 106 & 194 & --- & ---
  & 32  & 268 & --- & --- \\
\textsc{Self-Fix}
  & 115 & 185 & 9/194   & 4.6\%
  & 38  & 262 & 6/268   & 2.2\% \\
\textsc{DualFix}
  & 145 & 155 & 39/194  & 20.1\%
  & 63  & 237 & 31/268  & 11.6\% \\
\textsc{Iterative-DualFix}
  & \textbf{146} & \textbf{154} & \textbf{40/194}  & \textbf{20.6\%}
  & \textbf{73}  & \textbf{227} & \textbf{41/268}  & \textbf{15.3\%} \\

\midrule
\multicolumn{9}{l}{\textit{Codestral-22B}} \\
\midrule
Direct generation
  & 110 & 190 & --- & ---
  & 47  & 253 & --- & --- \\
\textsc{Self-Fix}
  & 125 & 175 & 15/190  & 7.9\%
  & 62  & 238 & 15/253  & 5.9\% \\
\textsc{DualFix}
  & 159 & 141 & 49/190  & 25.8\%
  & 91  & 209 & 44/253  & 17.4\% \\
\textsc{Iterative-DualFix}
  & \textbf{167} & \textbf{133} & \textbf{57/190}  & \textbf{30.0\%}
  & \textbf{101} & \textbf{199} & \textbf{54/253}  & \textbf{21.3\%} \\

\bottomrule
\end{tabular}
\end{table*}

Table~\ref{tab:main_results} compares DualFix against direct generation and Self-Fix across both benchmarks and generators. DualFix consistently outperforms the baseline in every setting. On Codestral-22B, DualFix fixes 25.8\% of all failures on LCB and 17.4\% on APPS. On Qwen-7B, it fixes 20.1\% on LCB and 11.6\% on APPS. Compared to Self-Fix, which relies solely on the generator's error signal and fixes at most 7.9\% of failures, DualFix fixes 3--5$\times$ more problems across all settings.

We see some differences among the fixes made by DualFix and the baseline (Self-fix): DualFix fixed 29, 17, 17, 34 cases that Self-fix did not for LCB–Qwen, LCB–Codestral, APPS–Qwen and APPS–Codestral, while Self-fix fixed 4, 7, 3, 5, with 5, 8, 3, 10 fixed by both. 

Additional error-repair rounds provide varying returns depending on the benchmark. On LCB, Iterative-DualFix yields only marginal gains over DualFix (20.1\% vs.\ 20.6\% for Qwen, 25.8\% vs.\ 30.0\% for Codestral), suggesting that most failures on these tasks are resolved in a single round. On APPS, the additional rounds provide a benefit (11.6\% vs.\ 15.3\% for Qwen, 17.4\% vs.\ 21.3\% for Codestral) producing richer error signals across iterations. 

Interestingly, we observe that in both cases, execution feedback has limits: in most of the failing cases, the error signal is not translated to any improvement since the LLMs cannot link the error feedback with the expected behavior. This results in some kind of a ``deadlock'', the LLMs make similar mistakes in multiple round despite the error feedback they get. The rule-based on the other side addresses these cases by rewriting the specification before any code is generated, sidestepping uninformative error signals entirely.

\begin{tcolorbox}[colback=blue!5, colframe=blue!40, boxrule=0.4pt, arc=2pt, left=4pt, right=4pt, top=2pt, bottom=2pt]

\textbf{Takeaway RQ1.} Combining the evolved rewriting rules with error feedback (DualFix) consistently outperforms error-only repair, turns to passing 3--5$\times$ more failures than Self-Fix. 

\end{tcolorbox}

\subsection{RQ2: Ablation study}

\begin{figure*}[t]
    \centering
    \includegraphics[width=0.70\linewidth]{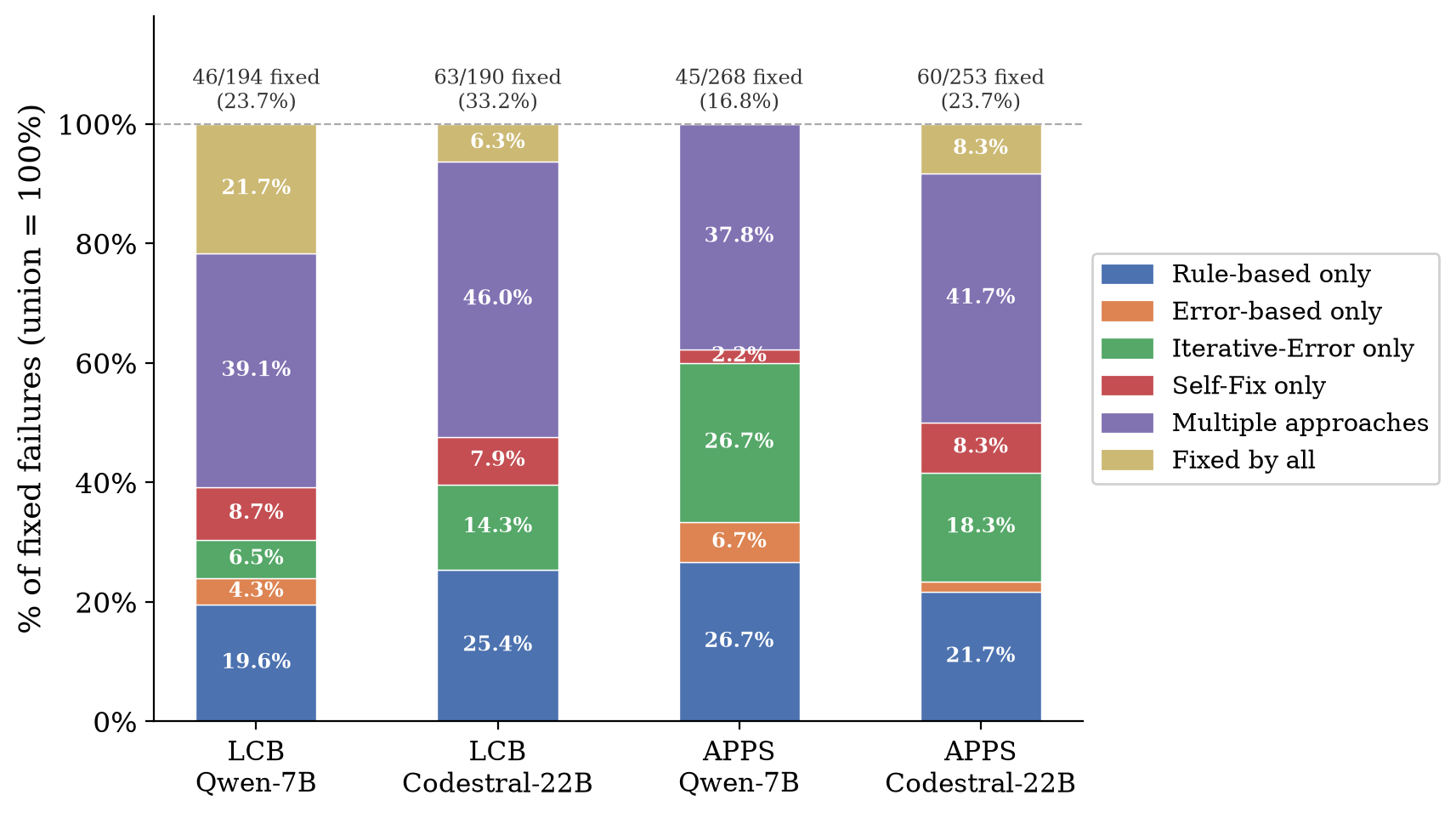}
    \caption{Overlap analysis of fixed failures across model-benchmark settings. Each bar shows the proportion of acceptance rate attributed to each approach alone, to multiple approaches, or to all. Rule-based rewriting and error-based repair fix largely disjoint failure sets.}
    \label{fig:overlap}
\end{figure*}

To isolate the contribution of each DualFix component, we evaluate Rule-Based rewriting, single-round Error-Based repair, and Iterative-Error ($\times$3) independently on all baseline failures. Table~\ref{tab:ablation} shows the results.

Rule-Based rewriting alone turns passing 11.9--21.6\% of failures 
on LCB and 8.2--13.8\% on APPS, consistently outperforming single-round Error-Based repair in three of four settings. On Codestral-22B, Rule-Based even surpasses Iterative-Error ($\times$3) on both benchmarks (21.6\% vs.\ 19.5\% acceptance rate on LCB, 13.8\% vs.\ 15.0\% on APPS), despite having no access to any runtime signal. This confirms that specification-level defects represent a substantial and distinct source of failures that error feedback alone cannot address.

Figure~\ref{fig:overlap} shows the overlap between the failure sets that turns to passing by each approach. Across all four settings, Rule-Based and error-based repair fix largely disjoint sets of problems: 4.5--8.4\% of baseline failures are fixed exclusively by Rule-Based rewriting, while only 0--1.1\% are fixed exclusively by single-round Error-Based repair. The low overlap explains why DualFix consistently outperforms either channel alone; the two strategies target fundamentally different defect classes, and composing them yields additive rather than redundant gains. Notably, 66.8--83.2\% of baseline failures remain unsolved by any approach, indicating significant room for future improvement.

\begin{tcolorbox}[colback=blue!5, colframe=blue!40, boxrule=0.4pt, arc=2pt, left=4pt, right=4pt, top=2pt, bottom=2pt]

\textbf{Takeaway RQ2.} Rule-Based rewriting and error-based repair fix largely disjoint failure sets. Rules alone match or exceed iterative error repair despite using no runtime signal, confirming that the way a prompt is worded can both improve or degrade downstream performance.
\end{tcolorbox}

\begin{table*}[t]
\centering
\small
\caption{Ablation study: Acceptance rate of each DualFix component 
on baseline failures.
\textsc{Rule-Based} = specification rewriting only.
\textsc{Error-Based} = single error-repair round only.
\textsc{Iterative-Error} = three error-repair rounds only.}
\label{tab:ablation}
\begin{tabular}{l  r r r r  r r r r}
\toprule
 &
  \multicolumn{4}{c}{\textbf{LCB} (300 problems)} &
  \multicolumn{4}{c}{\textbf{APPS} (300 problems)} \\
\cmidrule(lr){2-5}\cmidrule(lr){6-9}
\textbf{Component} &
  \textbf{Passing} & \textbf{Failing} & \textbf{Fixed} & \textbf{Rec.\,rate} &
  \textbf{Passing} & \textbf{Failing} & \textbf{Fixed} & \textbf{Rec.\,rate} \\

\midrule
\multicolumn{9}{l}{\textit{Qwen2.5-Coder-7B-Instruct}} \\
\midrule
Direct generation
  & 106 & 194 & --- & ---
  & 32  & 268 & --- & --- \\
\textsc{Rule-Based}
  & 129 & 171 & 23/194 & 11.9\%
  & 54  & 246 & 22/268 & 8.2\% \\
\textsc{Error-Based}
  & 133 & 167 & 27/194 & 13.9\%
  & 45  & 255 & 13/268 & 4.9\% \\
\textsc{Iterative-Error ($\times$3)}
  & 136 & 164 & 30/194 & 15.5\%
  & 60  & 240 & 28/268 & 10.4\% \\

\midrule
\multicolumn{9}{l}{\textit{Codestral-22B}} \\
\midrule
Direct generation
  & 110 & 190 & --- & ---
  & 47  & 253 & --- & --- \\
\textsc{Rule-Based}
  & 151 & 149 & 41/190 & 21.6\%
  & 82  & 218 & 35/253 & 13.8\% \\
\textsc{Error-Based}
  & 130 & 170 & 20/190 & 10.5\%
  & 71  & 229 & 24/253 & 9.5\% \\
\textsc{Iterative-Error ($\times$3)}
  & 147 & 153 & 37/190 & 19.5\%
  & 85  & 215 & 38/253 & 15.0\% \\

\bottomrule
\end{tabular}
\end{table*}

\subsection{RQ3: Rules generalization (transferability)}

\begin{table*}[t]
\centering
\small
\caption{Acceptance rate results for Claude Haiku~4.5 on LCB and APPS. 
Rules transferred zero-shot from Codestral-22B (no Claude-specific 
training). Rec.\,rate = Fixed / Failures.}
\label{tab:transfer_claude}
\begin{tabular}{l  r r r r  r r r r}
\toprule
 &
  \multicolumn{4}{c}{\textbf{LCB} (300 problems)} &
  \multicolumn{4}{c}{\textbf{APPS} (300 problems)} \\
\cmidrule(lr){2-5}\cmidrule(lr){6-9}
\textbf{Approach} &
  \textbf{Passing} & \textbf{Failing} & \textbf{Fixed} & \textbf{Rec.\,rate} &
  \textbf{Passing} & \textbf{Failing} & \textbf{Fixed} & \textbf{Rec.\,rate} \\
\midrule
Direct generation
  & 228 & 72  & ---    & ---
  & 192 & 108 & ---    & --- \\
\textsc{Self-Fix}
  & 248 & 52  & 20/72  & 27.8\%
  & 214 & 86  & 22/108 & 20.4\% \\
\textsc{Rule-Based}
  & 254 & 46  & 26/72  & 36.1\%
  & 222 & 78  & 30/108 & 27.8\% \\
\textsc{Error-Based}
  & 249 & 51  & 21/72  & 29.2\%
  & 207 & 93  & 15/108 & 13.9\% \\
\textsc{DualFix}
  & 260 & 40  & 32/72  & 44.4\%
  & 225 & 75  & 33/108 & 30.6\% \\
\textsc{Iterative-Error ($\times$3)}
  & 255 & 45  & 27/72  & 37.5\%
  & 220 & 80  & 28/108 & 25.9\% \\
\textsc{Iterative-DualFix}
  & \textbf{261} & \textbf{39} & \textbf{33/72}  & \textbf{45.8\%}
  & \textbf{231} & \textbf{69} & \textbf{39/108} & \textbf{36.1\%} \\
\bottomrule
\end{tabular}
\end{table*}

To evaluate generalization/transferability, we apply the rule set evolved on Codestral-22B directly to Claude Haiku~4.5 without any re-optimization. Table~\ref{tab:transfer_claude} reports the transferability results for both benchmarks.

The evolved rules consistently outperform all error-based baselines on both datasets. On LCB, Rule-Based rewriting alone turns passing 36.1\% of failures, outperforming Self-Fix (27.8\%) and Error-Based repair (29.2\%). On APPS, the gains are even larger: Rule-Based passes 27.8\% of failures, surpassing both Self-Fix (20.4\%) and Error-Based (13.9\%). The full  DualFix pipeline passes 45.8\% of failures on LCB and 36.1\% on APPS (33 and 39 problems, respectively). These results hold across two benchmarks with different problem formats, suggesting that the evolved rules are not overfit to the model or dataset they were optimized on.

To better understand the nature of the evolved rules and why they transfer, we analyze all rule sets produced across the four model-benchmark configurations (Table~\ref{tab:rule-comparison}). We observe three categories:  (i) four \emph{universal} rules that emerge independently in every setting, covering algorithmic terminology, math notation,  binary predicates, and function naming; (ii) \emph{benchmark-specific} rules that reflect the dominant challenge of each dataset, mathematical notation for LCB, procedural narratives for APPS; and (iii) \emph{model-specific} rules, such as backtick formatting for the smaller Qwen-7B. The universal core is present in every evolved rule set, which accounts for the successful zero-shot transfer: the rules that matter most are not tied to a particular model or benchmark but capture general specification clarity patterns.

\begin{tcolorbox}[colback=blue!5, colframe=blue!40, boxrule=0.4pt, 
  arc=2pt, left=4pt, right=4pt, top=2pt, bottom=2pt]
\textbf{Takeaway RQ3.} Rules evolved on one model transfer zero-shot to another (not used during rule evolution) model, outperforming its error-based baselines. The search independently converges on a shared core of specification clarity patterns across all configurations.
\end{tcolorbox}

\begin{table*}[t]
\centering
\caption{Cross-configuration rule analysis. $\checkmark$ = rule emerged during evolution. }
\label{tab:rule-comparison}
\small
\renewcommand{\arraystretch}{1.3}
\setlength{\tabcolsep}{5pt}
\begin{tabular}{p{1.2cm}lccccl}
\toprule
 & & \multicolumn{2}{c}{\textbf{LCB}} & \multicolumn{2}{c}{\textbf{APPS}} & \\
\cmidrule(lr){3-4} \cmidrule(lr){5-6}
\textbf{Category} & \textbf{Rule pattern} & Qwen & Codestral & Qwen & Codestral & \textbf{Interpretation} \\
\midrule
\multirow{4}{1.2cm}{\textit{Universal}}
 & Algorithmic terminology clarification  & $\checkmark$ & $\checkmark$ & $\checkmark$ & $\checkmark$ & \multirow{4}{4.8cm}{\raggedright Core semantic clarification patterns that generalize across all models and benchmarks.} \\
 & Math symbols $\rightarrow$ plain English           & $\checkmark$ & $\checkmark$ & $\checkmark$ & $\checkmark$ & \\
 & Binary predicate completion             & $\checkmark$ & $\checkmark$ & $\checkmark$ & $\checkmark$ & \\
 & Function name neutralization            & $\checkmark$ & $\checkmark$ & $\checkmark$ & $\checkmark$ & \\
\midrule
\multirow{4}{1.2cm}{\textit{LCB-specific}}
 & Modular arithmetic clarification        & & $\checkmark$ & & & \multirow{4}{4.8cm}{\raggedright LCB problems rely on heavy mathematical notation; rules target symbolic and exponent handling.} \\
 & Sequence notation $\rightarrow$ prose   & $\checkmark$ & & & & \\
 & Exponent expansion to decimal           & $\checkmark$ & & & & \\
 & Preserve subscript notation as-is       & & $\checkmark$ & & & \\
\midrule
\multirow{3}{1.2cm}{\textit{APPS-specific}}
 & Element-wise transform.\ preservation & & & $\checkmark$ & & \multirow{3}{4.8cm}{\raggedright APPS interview problems use procedural narratives; rules preserve or clarify operational sequences.} \\
 & Repeated-operation clarification        & & & & $\checkmark$ & \\
 & Periodic event timing                   & & & & $\checkmark$ & \\
\midrule
\multirow{2}{1.2cm}{\textit{Model-specific}}
 & Backtick formatting for variables       & $\checkmark$ & & $\checkmark$ & & \multirow{2}{4.8cm}{\raggedright Qwen-7B needs formatting support; Codestral-22B benefits from terminology generalization.} \\
 & Domain-specific term.\ $\rightarrow$ generic & & $\checkmark$ & & & \\ 
\bottomrule
\end{tabular}
\end{table*}

\section{Discussion}

\subsection{What Do the Evolved Rules Look Like?}

Figure~\ref{fig:rewrite-example} illustrates a representative fail-to-pass case. The original AtCoder problem uses raw LaTeX markup for constraints and states the task goal only in formal summation notation. Three of the seven evolved rules fire: R2 converts mathematical symbols to plain English, R1 restates the summation goal with a clarifying parenthetical, and R5 neutralizes the function signature. The rewritten version passes all tests without adding any new information. Notably, each rule targets a different surface pattern, notation, terminology, or naming, yet their combined effect resolves a specification-level failure that three rounds of error-based repair could not fix. This highlights a recurring pattern across our results: the rules do not solve the problem for the model; they remove surface-level ambiguity that prevents the model from recognizing a task it already knows how to solve.

\begin{figure*}[t]
\centering
\small
\begin{tcolorbox}[colback=gray!5, colframe=gray!70, boxrule=0.5pt, arc=2pt,
  title={\textbf{Fail-to-pass example: \texttt{abc379\_e} (AtCoder) --- Codestral-22B, LiveCodeBench}},
  fonttitle=\small]

\textbf{Original problem statement} (baseline: 0/2 tests passed) \\[4pt]
\textit{You are given a string S of length N consisting of digits from 1 through 9.
For each pair of integers (i,j) \textbackslash{} (1\textbackslash leq i\textbackslash leq j\textbackslash leq N), define f(i, j) as the value obtained by interpreting the substring of S from the i-th through the j-th character as a decimal integer. Find \textbackslash displaystyle \textbackslash sum\_\{i=1\}\^{}N \textbackslash sum\_\{j=i\}\^{}N f(i, j).}

\smallskip
\textit{Constraints: \texttt{1 \textbackslash leq N \textbackslash leq 2 \textbackslash times 10\^{}5}}

\tcblower

\textbf{Rewritten problem statement} (after rules: 2/2 tests passed) \\[4pt]
\textit{You are given a string S of length N consisting of digits from 1 through 9.
For each pair of integers (i,j) \textbf{(where $1 \leq i \leq j \leq N$)}, define f(i, j) as the value obtained by interpreting the substring of S from the i-th through the j-th character as a decimal integer. \textbf{Find the sum of f(i, j) for all valid pairs (i, j), which can be expressed as} $\displaystyle \sum_{i=1}^N \sum_{j=i}^N f(i, j)$.}

\smallskip
\textit{Constraints: \textbf{N is at least 1 and at most 200000}}

\tcblower

\textbf{Rules triggered:} \\[2pt]
\begin{tabular}{@{}lp{12.5cm}@{}}
\textit{Math symbols $\rightarrow$ plain English} &
Constraint \texttt{1 \textbackslash leq N \textbackslash leq 2 \textbackslash times 10\^{}5} rewritten as ``N is at least 1 and at most 200000''; LaTeX inequality notation replaced with Unicode symbols. \\[2pt]
\textit{Terminology clarification} &
Summation goal restated in natural language (``Find the sum of f(i, j) for all valid pairs'') before the formal expression. \\[2pt]
\textit{Domain-specific $\rightarrow$ generic} &
Raw LaTeX markup simplified to rendered mathematical notation.

\end{tabular} \\[6pt]
\textbf{Exact rules applied:} \\[3pt]
{\footnotesize
\textbf{R1.} \textsc{if} precise algorithmic or mathematical 
terminology appears, \textsc{then} add a short clarifying 
parenthetical. \textsc{except} when the phrase denotes a 
standard relation or order. \\[2pt]
\textbf{R2.} \textsc{if} constraints are written with 
mathematical symbols ($\leq$, $\geq$, \^{}), \textsc{then} 
rewrite in plain English preserving exact comparators and 
limits. \\[2pt]
\textbf{R5.} \textsc{if} the prompt includes an explicit 
required function name or signature, \textsc{then} rewrite 
it to a neutral, generic one. \textsc{except} when the 
signature is part of an online judge API.
}

\end{tcolorbox}
\vspace{-0.5em}
\caption{A fail-to-pass example on Codestral-22B. The original problem uses LaTeX markup and presents the task goal only in formal notation. Three evolved rules triggered: constraints were expanded to plain English, the summation goal was clarified in natural language, and LaTeX markup was simplified. No new information was added. The rewritten version passes all test cases.}
\label{fig:rewrite-example}
\end{figure*}

\subsection{Evolution of natural language transformation rules}

\begin{figure}
    \centering
    \includegraphics[width=\linewidth]{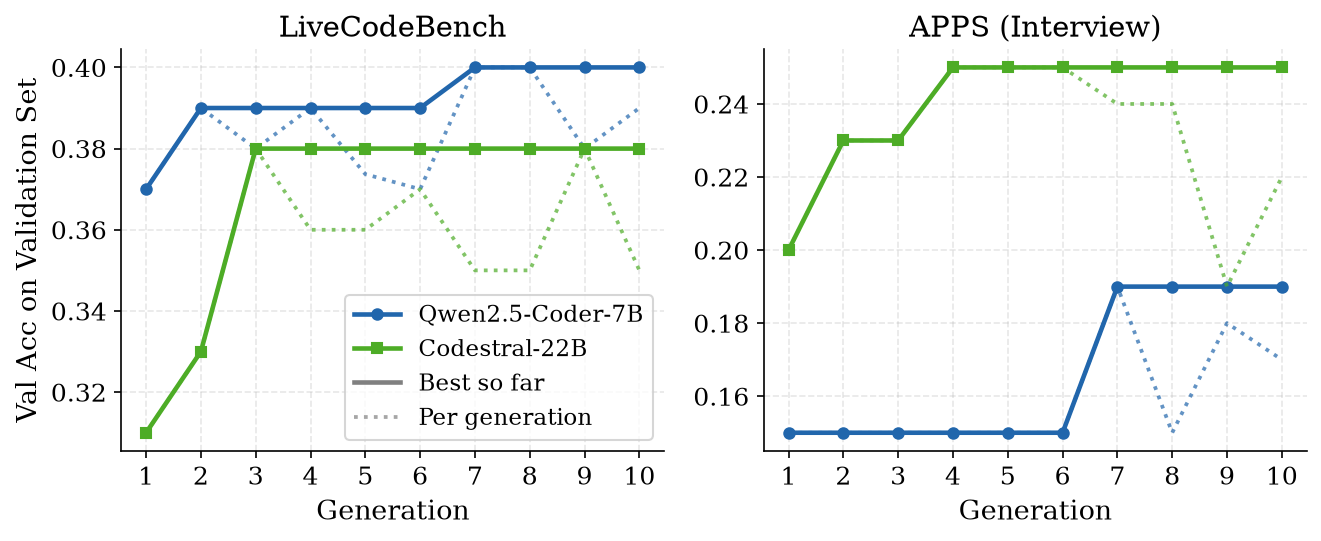}
    \vspace{-0.5em}
    \caption{Validation accuracy of the best rule set across evolutionary generations on LiveCodeBench (left) and APPS (right) for both generators.}
    \label{fig:convergence}
\end{figure}

Figure~\ref{fig:convergence} shows the validation accuracy of the best rule set across evolutionary generations. On both benchmarks, the search improves steadily over the first five generations before plateauing, confirming that the genetic loop discovers increasingly effective rules. At the same time, these results show that the approach converges fast, showcasing that it may require more stochastic mutations or a significant higher number of additional generations to further improve its performance. Nevertheless, it is interesting to see that the approach provides actionable results in a relatively low number of generations.

\subsection{Which Mutation Operators Drive the Search?}

Table~\ref{tab:mutations} summarizes the accepted mutations across all four configurations. Reflect and Merge each account for 32 accepted candidates, followed by Delete with 28, for an overall acceptance rate of 52\%. However, the three operators play distinct roles.

Reflect is the primary driver of rule \emph{evolution}: it produced the final selected rule set in both LCB configurations, confirming that LLM-guided editing informed by categorized execution feedback is the most effective mechanism for refining rules. Deletion produced the final selected rule set in both APPS configurations, indicating that interview style problems are particularly sensitive to over-specification; removing rules that introduce regressions or interfere with other rules can be more beneficial than adding new ones. Merge achieves a 100\% acceptance rate because it bypasses the gate batch and is admitted whenever the merged candidate claims at least one Pareto win. Despite this, no merge candidate was selected as the final best rule set in any configuration, suggesting that combining complementary patterns into a single rule set dilutes the specificity that makes individual candidates strong on unseen problems. Rule diversity is thus better maintained through the Pareto frontier, which keeps candidates in parallel, than through direct combination.

\begin{table}[t]
\centering
\small
\caption{Accepted mutations by operator. $\star$ = produced 
the final selected rule set. Acc.\,rate = total accepted / 
total generated across all operators.}
\label{tab:mutations}
\begin{tabular}{lrrrr}
\toprule
\textbf{Configuration} & \textbf{Reflect} & \textbf{Delete} & \textbf{Merge} & \textbf{Acc.\,rate} \\
\midrule
LCB - Qwen       & 8$\star$  & 9        & 8 & 25/44 (57\%) \\
LCB - Codestral  & 11$\star$ & 8        & 8 & 27/44 (61\%) \\
APPS - Qwen      & 6         & 8$\star$ & 8 & 22/44 (50\%) \\
APPS - Codestral & 7         & 3$\star$ & 8 & 18/44 (41\%) \\
\midrule
\textbf{Total}  & \textbf{32} & \textbf{28} & \textbf{32} & \textbf{92/176 (52\%)} \\
\bottomrule
\end{tabular}
\end{table}

\section{Threats to validity }
\textit{Internal validity.} The seed rules used to initialize the evolutionary search may bias the final rule sets. To assess this, we conducted an additional experiment in which the population was initialized with an empty rule set. The search independently rediscovered core patterns such as the math-symbols-to-plain-English rule, confirming that the evolutionary loop can converge on effective transformations without manual initialization. However, the empty-seed rules improved training accuracy but did not generalize to the test set, suggesting that the seed rules provide useful inductive bias that accelerates convergence toward generalizable patterns. The Rewriter (GPT-4o-mini) and Mutator (GPT-5-mini) are external dependencies; different models in these roles may produce different rule sets. However, the evolved rules are standalone text artifacts that, once discovered, can be applied by any instruction-following model. The data split relies on a fixed random seed ($s$=42); different splits could affect which rules survive, though the universal core (Table~\ref{tab:rule-comparison}) emerges consistently across four independent runs. \\

\textit{External validity.} We evaluate on two benchmarks (LiveCodeBench and APPS) and three generators of varying size and architecture (Qwen-7B, Codestral-22B, Claude Haiku~4.5). While these cover diverse problem formats, functional and stdin/stdout, competitive programming, and interview-level, results may not generalize to other programming languages beyond Python, to non-competitive problem styles such as open-ended software engineering tasks, or to significantly larger models where prompt sensitivity patterns may differ. APPS is filtered to the interview tier; competition-level problems involving deep algorithmic reasoning may respond differently to specification rewriting.

\textit{Construct validity.} We report pass@1 under greedy decoding (temperature~$=0$) as the sole metric, which does not capture partial correctness or sampling-based evaluation. Test suites may be incomplete, and APPS test cases are capped at 10 per problem. However, test-suite overfitting is structurally unlikely for either repair channel. The error signal exposes only the input/output of the first failing test case; the Rewriter never sees the full test suite. The model must therefore infer the general bug from a single I/O pair, rather than memorizing multiple specific cases. Rule-based repair receives no test information at all, making it immune by construction.  The 30-second timeout may conflate slow solutions with specification defects. Finally, while rules are constrained to never add new information, we rely on the Rewriter to enforce this; subtle semantic shifts during rewriting cannot be fully ruled out.
\section{Conclusion}

LLMs are sensitive to prompt formulation, and this sensitivity extends to code generation: minor variations in how a programming task is worded can turn a passing solution into a failing one. Existing repair approaches focus on execution feedback, yet our results show that 10 - 19\% of failures resist multiple rounds of error-based repair, suggesting that at least some of these failures may stem from how the specification is worded rather than from the implementation itself.

We propose DualFix, a staged pipeline that combines two complementary repair channels: execution-feedback rewriting and error-agnostic specification rewriting driven by evolved transformation rules. The rules are discovered by RuleEvol, a search-based framework that uses a genetic loop to evolve compact, reusable IF-THEN-EXCEPT patterns that clarify task descriptions without altering their semantics.

Our evaluation on LiveCodeBench and APPS across three generators shows that DualFix consistently outperforms all error-only baselines, recovering up to 30\% of baseline failures and 3--5$\times$ more problems than Self-Fix. The ablation study reveals that the two channels fix largely disjoint failure sets, with rules alone matching or exceeding iterative error repair despite using no runtime signal. The evolved rules transfer zero-shot to an unseen model, and a shared core of four specification clarity patterns emerges independently across all configurations. These findings suggest that how a task is worded can be as much a source of failure as how the code is written, and that fixing the specification complements rather than duplicates fixing the code.

Future work includes designing transformation rules that act as a general-purpose preprocessing filter, applied to all prompts before code generation, not only to failing cases but to every task description, potentially preventing failures before they occur rather than repairing them afterward. Finally, characterizing the 66--83\% of failures that no current approach can reach may partly reflect inherent model capability limits or reveal fundamental limits of specification-level repair, and point toward new strategies altogether.


\clearpage
\bibliographystyle{IEEEtranS}
\bibliography{literature}

\end{document}